\documentclass[showpacs,amsmath,
twocolumn,
aps,prl]{revtex4}
\usepackage{graphicx}
\usepackage{dcolumn}

\begin{document}

\title{All-electrical measurement of spin injection in a magnetic $p$-$n$ junction diode}
\author{Peifeng Chen}
\author{Juergen Moser}
\author{Philipp Kotissek}
\author{Janusz Sadowski}
\author{Marcus Zenger}
\author{Dieter Weiss}
\author{Werner Wegscheider}
\affiliation{Institut f\"ur Experimentelle und Angewandte Physik, Universit\"at Regensburg, 93040 Regensburg, Germany.}
\date{\today}

\begin{abstract}
Magnetic $p$-$n$ junction diodes are fabricated to investigate spin-polarized electron transport.  The injection of spin-polarized electrons in a semiconductor is achieved by driving a current from a ferromagnetic injector (Fe), into a bulk semiconductor ($n$-GaAs) via Schottky contact.  For detection, a diluted magnetic semiconductor ($p$-GaMnAs) layer is used.  Clear magnetoresistance was observed only when a high forward bias was applied across the $p$-$n$ junction.  
\end{abstract}

\maketitle

Spin electronics has attracted growing interest in recent years, because it is envisioned that the utility of the spin and charge properties of the electron would open new perspectives to semiconductor device tech\-nology~\cite{Wolf:2001,Prinz:1998,Johnson:1993}.  One essential requirement for the spintronic devices is the efficient injection of spin-polarized carriers.  Spin injection from a ferromagnetic metal into a semiconductor is attractive, because ferromagnetic metals such as Fe and Co have a relatively high Curie temperature.  Spin LED structures provide a way to study the spin injection~\cite{Zhu:2001}.  The spin polarized carriers injected from the ferromagnetic metals radiatively recombine in the semiconductor emitting circularly polarized light.  It was found that Schottky or tunneling contacts between a metallic ferromagnet and a semiconductor can overcome the conductance mismatch obstacle and show carrier polarizations up to thirty percent~\cite{Erve:2004,Hanbicki:2002}.  

However, from a device point of view, a major breakthrough still would be to have all electronic devices.  Recently, magnetic $p$-$n$ junction diodes have been proposed and theoretically analyzed~\cite{Zutic:2002,fabian:2002}.  These devices whose electronic properties depend on the spin polarization of the carriers can offer opportunities to study the effective spin injection.  In this paper, we demonstrate the fabrication of a novel magnetic $p$-$n$ junction diode, in which the spin injection between ferromagnetic metals and semiconductors is measured all-electrically.  

\begin{figure}[htbp]
  \includegraphics[width=7cm]{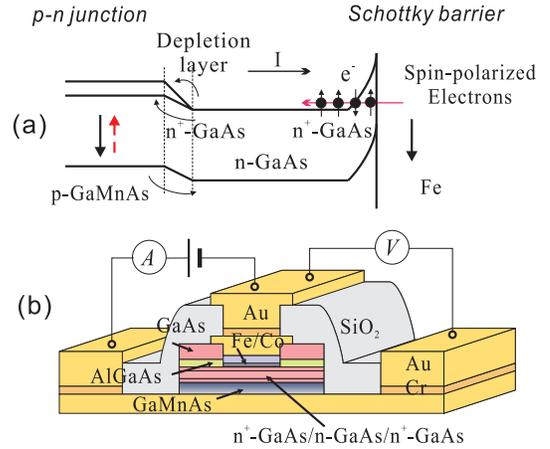}
  \caption{
     (a) Band-energy schemes for the magnetic $p$-$n$ junction with a Fe/GaAs Schottky barrier. The $p$ region (left) is magnetic GaMnAs layer, indicated by the spin splitting of the conduction band. Under applied forward bias, the spin-polarized electrons (solid circles) are injected from Fe (right) to the $n$-GaAs region (middle) and extracted across the depletion layer into the $p$ region.  Up and down arrows indicate the magnetizations of the two electrodes.  (b) The cross section of the device geometry with four-probe measurements.     
     }
  \label{fig:structure}
\end{figure}

A sketch of the band diagram for such a magnetic/nonmagnetic semiconductor $p$-$n$ junction in contact with a ferromagnetic metal is presented in Fig.~\ref{fig:structure}(a).  The device performs as follows: a positive bias is applied between the $p$-GaMnAs and the ferromagnetic Fe layers. This places the magnetic $p$-$n$ junction in forward bias and the Fe/GaAs Schottky barrier in reverse bias.  Consequently the spin-polarized electrons are injected from Fe into the bulk $n$-GaAs via a Schottky contact.  Afterwords, the spin-polarized electrons are extracted from the  $n$-GaAs across the depletion layer into the $p$-GaMnAs.  If the relative magnetizations of the two magnetic electrodes are changed from parallel to antiparallel, the magnetic $p$-$n$ junction diode should display a GMR-like effect.  
      
The preparation of the hybrid structure is started from the semiconductor heterostructure, which was grown on a semi-insulating GaAs substrate by molecular beam epitaxy at growth temperature of $630\,^{\circ}\mathrm{C}$.  It has the following layer sequence: 300nm GaAs buffer layer/ 300nm AlAs-GaAs super\-lattice/ 100nm GaAs/ 50nm Al$_{0.72}$Ga$_{0.18}$As/ 15nm $n^+$-GaAs($3\times10^{18}$cm$^{-3}$)/ 50nm $n$-GaAs($1\times10^{16}$cm$^{-3}$)/ 10nm $n^+$-GaAs($3\times10^{18}$cm$^{-3}$)/ 60nm $p$-Ga$_{0.94}$Mn$_{0.06}$As.  In the sample structure, the 50nm $n$-GaAs is used as transport region for spin-polarized electrons.  The 10nm $n^+$-GaAs layer with a Si doping density of $3\times10^{18}$cm$^{-3}$ leads to a smaller depletion region between the $p$-GaMnAs and bulk $n$-GaAs layer.  The other 15nm $n^+$-GaAs is used to control the Schottky barrier interface resistivity to overcome the conductance mismatch between Fe and GaAs~\cite{Schmidt:2000,Rashba:2000,Fert:2001}.  The GaMnAs layer was grown at a growth temperature of $250\,^{\circ}\mathrm{C}$ and shows a Curie temperature of 65K. 

In order to realize the device, we use the epoxy bonding and stop-etching technique (EBASE)~\cite{Kreuzer:2002,Zenger:2004,Weck:1996}, which relies on the highly selective etching of GaAs and Al$_x$Ga$_{1-x}$As by suitable wet chemical etchant.  The fabrication steps of the sample involve conventional optical lithography and lift-off procedure.  A 100nm Au film deposited on the GaMnAs layer is used as contact to the soft magnetic electrode.  Before depositing the second contact, the original sample is inverted and epoxy bonded onto a new semi-insulating host substrate and cured at $80\,^{\circ}\mathrm{C}$ for 4 hours.  The original substrate and the 300nm thick superlattice are removed then. After the 100nm GaAs and 50nm AlGaAs layers are selectively etched by the citric acid and 1\% HF respectively, the sample is transferred to the sputtering system immediately.  The GaAs semiconductor surface is treated with H$^+$ plasma to remove the oxide layer, then a 12nm Fe layer and a 50nm Co magnetic pinning layer~\cite{Vincent:2003} are deposited on the $n$-GaAs layer as a hard magnetic electrode. Finally, photolithographic definition of a mesa structure provides access to the bottom voltage probes and the thick SiO$_2$ film deposition is used for electrical isolation.  The cross section of the whole device for four-probe measurements is shown in Fig.~\ref{fig:structure}(b).  Due to the extreme selectivity of HF ($\geq10^7$), the transport length of bulk $n$-GaAs was precisely defined by MBE growth only~\cite{Yabl:1987}.  The electric and magnetotransport characterizations of the ferromagnet based magnetic/nonmagnetic p-n junction were carried out at room temperature and at 4.2K employing an HP Semiconductor Analyzer 4155A.  The sample was mounted in a $^4$He cryostat with a superconducting coil and the magnetic field was aligned in the plane of the hybrid structure.  
\begin{figure}[htbp]
  \includegraphics[width=7cm]{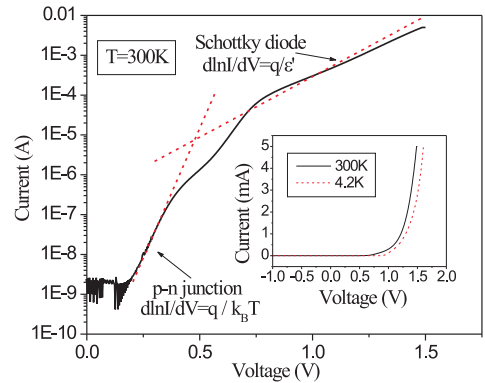}
  \caption{
  Logarithmic plot of forward I-V characteristic of the magnetic $p$-$n$ junction diode at room temperature.  The dashed lines represent the theoretical slopes of the curve, indicating the I-V characteristic dominated by the $p$-$n$ junction and Schottky diode respectively.  The inset shows the I-V curves of the device at 4.2K (dashed line) and room temperature (solid line).  
  }
  \label{fig:IV}
\end{figure}

The I-V  curves of the magnetic $p$-$n$ junction diode measured at room temperature and 4.2K are shown in the inset in Fig.~\ref{fig:IV}.  If we look closer at the logarithmic plot of the current vs applied voltage, the different slopes of the curve can be found, see Fig.~\ref{fig:IV}.  The device studied here can be treated as a stack of a $p$-$n$ junction and a Schottky diode.  For the $p$-$n$ junction, the current can be expressed as $J=J_{s1}[\exp(qV_1/k_BT)-1]$, where $J_{s1}$ is the saturation current density, $V_1$ is the bias for the $p$-$n$ junction, $q$ is the magnitude of electronic charge, $k_B$ is Boltzmann's constant and $T$ is the absolute temperature.  
\begin{figure}[htbp]
  \includegraphics[width=8.5cm]{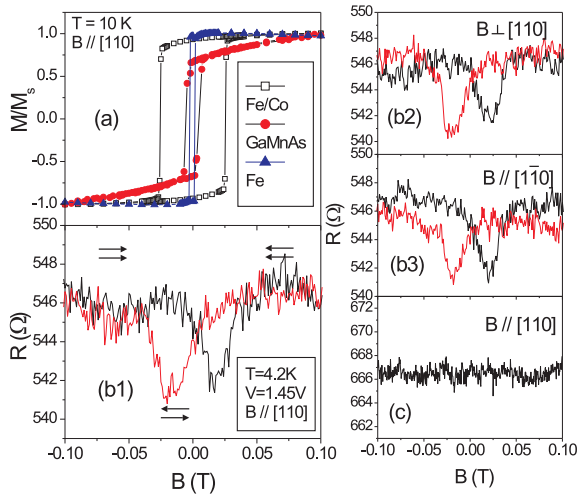}
  \caption{
  (a) SQUID magnetic measurements of the GaMnAs, Fe and Fe/Co films at 10K.  (b) Magnetoresistance of the device ($p$-GaMnAs/$n$-GaAs/Fe/Co) plotted as a function of magnetic field in the plane.   The in-plane magnetic field is along [110] (b1), perpendicular to [110] (b2) and along [1\={1}0] (b3).  The alignment of the electrodes is indicated by arrows in figure (b1).  (c) Magnetoresistance curve of the reference sample ($p$-GaMnAs/$n$-GaAs/Fe).  Constant resistance is observed.      
    }
  \label{fig:HysMR}
\end{figure}
Since a heavily doped $n^+$-GaAs with a doping density of $N_d=3\times10^{18}$cm$^{-3}$ was used for the Schottky contact, the tunneling of electrons through the barrier plays an important role in the transport process.  For the Schottky diode under reverse bias, the current density can be expressed as $J=J_{s2}\exp(qV_2/\varepsilon')$, where $J_{s2}$ is the saturation current density and  $V_2$ is bias for the Schottky diode~\cite{Pado:1966}.  From the definitions: $\varepsilon'=E_{00}[(E_{00}/k_BT-\tanh{E_{00}/k_Bt})]^{-1}$, $E_{00}=(qh/4\pi)[N_d/m^*\epsilon]$, where $m^*$  is the  effective mass and $\epsilon$ is the dielectric constant, we obtain $\varepsilon'=78.5$meV.  Since $\varepsilon'$ is three times as large as $k_BT$ at room temperature, the voltage drop over the Schottky barrier increases faster than the voltage drop over the $p$-$n$ junction, for increasing current in the device.  At low voltage, the resistance of the Schottky barrier is much lower than that of the $p$-$n$ junction.  Consequently, the I-V characteristic is dominated by the $p$-$n$ junction with a slope equal to $q/k_BT$.  When the voltage is increased, the resistance of the Schottky barrier becomes comparable to the $p$-$n$ junction and cannot be neglected, and at high voltage, the slope equals to $q/\varepsilon'$ as shown in Fig.~\ref{fig:IV}. 

Fig.~\ref{fig:HysMR}(a) shows the magnetic hysteresis loops at 10K for the GaMnAs (60nm), Fe (22nm) and Fe (12nm)/Co (50nm) layers.  Using a Co layer to magnetically bias the Fe film, the coercivity of the 12nm Fe layer is 30mT, while the GaMnAs layer shows a coercivity of 3mT.  In the magnetic field range between these coercivities, the ferromagnets' magnetization can be switched to the  anti\-parallel state.  Fig.~\ref{fig:HysMR}(b1) shows the magnetoresistance curve of the device measured at 4.2K, the applied magnetic field is in the plane and parallel to the [110] direction of GaMnAs. The negative magnetoresistance curve coincides reasonably well with the distinct coercive fields of the magnetization curves of GaMnAs and Fe/Co layers.  With forward applied bias of 1450mV, a negative magnetoresistance of 1.02\% is found.  The detailed mechanism of the negative magnetoresistance is still not clear.  It is probably due to the antiferromagnetic $s$-$d$ exchange in GaMnAs~\cite{Myers:2005}.  For parallel magnetic configuration, the barrier for crossing the depletion layer is larger for the spin up electrons (majority).  Therefore, we find the negative magnetoresistance in our experiments.  

In order to exclude the tunneling anisotropic magnetoresistance (TAMR) effect in our results~\cite{Ruester:2005}, the angle dependence of the magnetoresistance was studied.  In the measurements, the angle of the in-plane magnetic field was changed with respect to the [110] crystallographic direction.  Two magnetoresistance curves with the magnetic field perpendicular to the [110] and along [1\={1}0] are shown in Fig.~\ref{fig:HysMR}(b2) and (b3).  Since the magnetoresistance does not change for different in-plane directions, we assume that the TAMR plays no important role here.  As another test of our results, measurements were made on a reference sample with a 22nm Fe layer in place of the 12nm Fe/50nm Co magnetic electrode. As shown in Fig.~\ref{fig:HysMR}(a), the Fe layer without Co pinning has a similar coercivity than the GaMnAs layer.  Thus, the magnetic configurations of these two electrodes can not switch from parallel to antiparallel.  The magnetoresistance is shown in Fig.~\ref{fig:HysMR}(c) and the value of the resistance remains constant.  This measurement of the reference sample gives further validation of our results. 
\begin{figure}[htbp]
  \includegraphics[width=6cm]{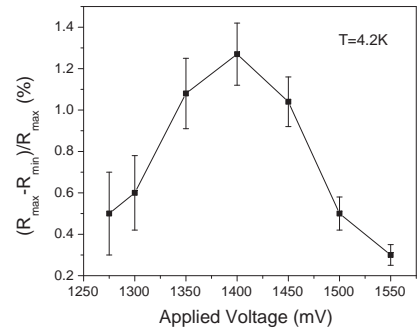}
  \caption{
  Magnetoresistance signal versus voltage.  The solid line is a guide to the eye.              
    }
  \label{fig:MRV}
\end{figure}

The bias voltage dependence of the magnetoresistance was also studied and the results are shown in Fig.~\ref{fig:MRV}.  We find the magnetoresistance can only be found with high forward bias on the device.  Theoretical analysis of the magnetic $p$-$n$ junction shows that there is no spin injection at small biases, because the injection of polarized carriers is still smaller than the equilibrium carrier density.  Typically the bias should be above 1V, therefore spin-polarized carriers can be injected across the depletion layer~\cite{fabian:2002}.   Our experimental results agree very well with the theory.  Furthermore, we also find a peak of the negative magnetoresistance signal of 1.27\% at 1400mV bias.  This effect can be explained by the interface resistance which was analyzed by Fert~\cite{Fert:2001}. In his model, the highest magnetoresistance is obtained in the limit $r_N(t_N/l^N_{sf})\ll r_b^*\ll r_N(l^N_{sf}/t_N)$, where $r_N$ is the product of the semiconductor resistivity and the spin diffusion length ($l^N_{sf}$), $t_N$ is the semiconductor transport length and $r_b^*$ is the interface resistance.  Since the resistivities of GaMnAs and GaAs are almost the same in our device, the interface resistivity is dominated by the Schottky barrier between Fe and GaAs.  When the voltage applied on the Fe/GaAs Schottky barrier varies, the interface resistivity is increased or decreased accordingly, hence the magnetoreistance reaches a maximum value in between.   

In conclusion, we have fabricated a novel magnetic $p$-$n$ junction diode, in which the spin injection from ferromagnetic metals to semiconductors can be measured all-electrically.  Our study shows the spin-polarized electrons can only be injected when a high forward bias is applied on the $p$-$n$ junction, which agrees well with the theoretical prediction.  Furthermore, the bias dependence of the magnetoresistance has also been discussed.   

\begin{acknowledgments}
The authors wish to thank Matthias Sperl for the SQUID measurements.  One of the authors (Peifeng Chen) would like to thank J.~Fabian and Shidong Wang for fruitful discussions.
\end{acknowledgments}


\end{document}